\documentclass[conference, 9pt, hidelinks]{IEEEtran}
\usepackage[nocompress]{cite}
\usepackage{graphicx}
\usepackage[T1]{fontenc}
\usepackage[tt=false, type1=true]{libertine}
\usepackage[varqu]{zi4}
\usepackage[libertine]{newtxmath}

\usepackage{amsmath}
\usepackage{comment}
\usepackage[table]{xcolor}
\usepackage{tabularx}
\usepackage{multirow, multicol}
\newcolumntype{Y}{>{\centering\arraybackslash}X} 
\usepackage{tabularray}
\usepackage{makecell}
\usepackage{array}
\usepackage{xspace}
\usepackage[font=small]{caption}
\usepackage{orcidlink}
\usepackage{nicefrac}
\usepackage{siunitx}
\usepackage{float}
\usepackage{enumitem}
\usepackage{subcaption}
\usepackage{relsize}
\usepackage{adjustbox}
\usepackage{soul}
\usepackage{amsthm}
\usepackage{mathtools}
\usepackage{etoolbox}
\usepackage{titlesec}
\usepackage{nicefrac}
\usepackage{diagbox}
\usepackage[hang, flushmargin]{footmisc}
\usepackage{algorithm}
\usepackage[noend]{algpseudocode}

\makeatother

\def\sothat{\,\lvert\,}
\newcommand\abs[1]{\lvert#1\rvert}

\newcommand{\kk}{\text{\scriptsize\sf k}\xspace}
\newcommand{\MM}{\text{\scriptsize\sf M}\xspace}

\DeclareMathOperator*{\argmax}{argmax}

\newcommand{\pmslash}{%
  \mathbin{%
    \vcenter{\hbox{%
      \setlength{\unitlength}{1em}%
      \begin{picture}(1,1)
        \put(0.10,0.55){\scriptsize$+$}
        \put(0.50,0.15){\scriptsize$-$}
        \put(0.50,0.50){\makebox(0,0){\rotatebox{-45}{\scriptsize$|$}}}
      \end{picture}
    }}%
  }\!\!%
}

\def\hedge{\mbox{h\!\!\;-\!\!\;edge}\xspace}
\def\hgraph{\mbox{h\!\!\;-\!\!\;graph}\xspace}
\def\hedges{\mbox{h\!\!\;-\!\!\;edges}\xspace}
\def\hgraphs{\mbox{h\!\!\;-\!\!\;graphs}\xspace}

\algnewcommand\ProcedureLine[1]{\item[\textbf{Procedure:}] #1}

\newcommand{\onesquishedline}[1]{%
  \noindent\hbox to \linewidth{%
    \spaceskip=0pt plus 1fil minus 1fil #1%
  }%
}

\newskip\paragraphbreak
\titleformat{\subsubsection}[runin]
  {\itshape} 
  {\thesubsubsection.}{0.5em}{}[: \quad] 

\titlespacing*\section{0pt}{*2}{*1}
\titlespacing*\subsection{0pt}{*2}{*1}
\titlespacing{\subsubsection}{0pt}{4pt}{0pt} 

\expandafter\def\expandafter\normalsize\expandafter{%
    \normalsize
    \setlength\abovedisplayskip{5pt} 
    \setlength\belowdisplayskip{5pt} 
    \setlength\abovedisplayshortskip{3pt} 
    \setlength\belowdisplayshortskip{3pt} 
}

\newlength{\itemizepadding}
\setlist[itemize]{topsep=\itemizepadding, leftmargin=3em} 
\setlist[enumerate]{topsep=\itemizepadding, leftmargin=3em} 

\makeatletter
\renewcommand{\title}[1]{\gdef\@title{#1\vspace{-9.5pt}}}
\makeatother

\renewcommand{\IEEEbibitemsep}{0pt plus 0.5pt}
\makeatletter
\renewcommand{\@biblabel}[1]{[#1]\hfill} 
\IEEEtriggercmd{\reset@font\normalfont\fontsize{7pt}{8pt}\selectfont} 
\IEEEtriggeratref{1}
\makeatother

\usepackage{nicematrix, tikz, calc, relsize}
\usetikzlibrary{calc}

\newcommand{\slantedssnstable}{
\begin{table}[t]
    \vspace{13pt}
    \centering
    \resizebox{\columnwidth}{!}{%
        \begin{minipage}{1.25\columnwidth}
            \centering
            \setlength{\tabcolsep}{0pt}
            \begin{NiceTabularX}{\textwidth}{X[1.4,c,m] *{10}{X[0.9,c,m]}}
                & & & & & & & & & & \\
                
                $\abs{N}$
                & 20\kk & 110\kk & 216\kk & 302\kk & 14\kk & 208\kk & 194\kk & $16\kk$ & $64\kk$ & $256\kk$ \\
                
                $\sum_{e \in E} \abs{e}$
                & 766\kk & 23\MM & 90\MM & 256\MM & 875\kk & 145\MM & 133\MM & 2.1\MM & 12.6\MM & 67.4\MM \\
                
                $\text{avg}_{e \in E} \abs{e}$
                & 37.3 & 210.3 & 417.2 & 848.1 & 63.2 & 696.2 & 688.3 & 128 & 192 & 256 \\

                \textbf{$\Omega, \, \Delta$}
                & ${\scriptstyle 2}^{10}\!\!, {\scriptstyle 2}^{12}$ & ${\scriptstyle 2}^{10}\!\!, {\scriptstyle 2}^{12}$
                & ${\scriptstyle 2}^{12}\!\!, {\scriptstyle 2}^{16}$ & ${\scriptstyle 2}^{12}\!\!, {\scriptstyle 2}^{16}$
                & ${\scriptstyle 2}^{10}\!\!, {\scriptstyle 2}^{12}$ & ${\scriptstyle 2}^{12}\!\!, {\scriptstyle 2}^{16}$
                & ${\scriptstyle 2}^{12}\!\!, {\scriptstyle 2}^{16}$ & ${\scriptstyle 2}^{10}\!\!, {\scriptstyle 2}^{12}$
                & ${\scriptstyle 2}^{10}\!\!, {\scriptstyle 2}^{12}$ & ${\scriptstyle 2}^{10}\!\!, {\scriptstyle 2}^{12}$ \\

                \CodeAfter
                \begin{tikzpicture}
                    \foreach \j in {1,2,...,12}
                        \draw (2-|\j) -- (6-|\j);
                
                    \foreach \i in {2,...,11}
                        \draw (2-|\i) -- ++(155:1.5cm); 
                
                    \foreach \label [count=\i from 1] in {
                        \textsmaller[1]{16\kk-model}, \textsmaller[1]{64\kk-model}, \textsmaller[1]{256\kk-model}, \textsmaller[1]{1\MM-model}, 
                        \textsmaller[1]{lenet}, \textsmaller[1]{alexnet}, \textsmaller[1]{\smash[b]{vgg11}},
                        \textsmaller[1]{16\kk-rand}, \textsmaller[1]{64\kk-rand}, \textsmaller[1]{256\kk-rand}
                    } {                            
                        \node[rotate=-25, anchor=south east, inner sep=-1pt] 
                            at ($ (2-|\inteval{\i+1}) + (5pt,3pt) $) {\label};
                    }
                    
                    \draw (2-|12) -- ++(155:1.5cm);

                    \draw (2-|2) -- (2-|12);
                \end{tikzpicture}
            \end{NiceTabularX}
        \end{minipage}%
    }
    \vspace{-2pt}
    \caption{Spiking neural networks used in the experiments.}
    \vspace{-4pt}
    \label{tab:snns}
\end{table}
}

\usepackage{booktabs}

\title{Incidence Constraints in Hypergraph Partitioning on GPU}

\author{
    Marco Ronzani \orcidlink{0009-0002-8485-0717}\,, Cristina Silvano \orcidlink{0000-0003-1668-0883}~\IEEEmembership{Fellow,~IEEE},
    \;DEIB, Politecnico di Milano, Italy\vspace{-18pt}
    \thanks{
        Manuscript received DD Month YYYY; revised DD Month YYYY; accepted DD Month YYYY. Date of publication DD Month YYYY; date of current version DD Month YYYY. (Corresponding author: Marco Ronzani.)
    }
    \thanks{
        The authors are with the Department of Electronics, Information and Bioengineering, Politecnico di Milano, Via Giuseppe Ponzio 34, 20133 Milano, Italy (e-mail: \url{marco.ronzani@polimi.it}; \url{cristina.silvano@polimi.it}).
    }
}

\begin{document}

\maketitle

\begin{abstract}
    Hypergraph partitioning is a pervasive NP-hard problem, and accelerating its computation on GPU can both slice time-to-solution and raise quality of results.
    In this work, we implement a multi-level hypergraph partitioning algorithm on GPU targeting a specific set of problem constraints: bounded per-partition size and distinct inbound hyperedges.
    Manipulating hypergraphs requires long orders of nested iterations, and enforcing these constraints introduces further set operations amidst them.
    Hence, we design algorithms around our problem's specifics, materializing the hypergraph’s incidence structure in memory and exploiting set sparsity.
    Our results show competitive speedups as high as $\mathbf{940\times}$ and $\mathbf{2}$-$\mathbf{26\%}$ better results in connectivity over a sequential multi-level partitioner.%
\end{abstract}

\begin{IEEEkeywords}
Hypergraph partitioning, GPU implementation, incidence constraint, size constraint.
\end{IEEEkeywords}

\section{Introduction}\label{sec:intro}

Hypergraph partitioning is a widely occurring problem in computer science.
Being notoriously NP-hard, it is typically addressed via heuristics \cite{AdvancesInHypergraphPartitioning}.
Hence, as problem size grows, more and more quality of results must be traded to retain a feasible time-to-solution.
For this reason, the efficient, massively parallel implementation of hypergraph partitioning algorithms holds the potential for time savings and performance improvements across many fields.
However, with the naturally sparse and unbalanced structure of hypergraphs, such algorithms are far from trivially parallelizable \cite{AdvancesInHypergraphPartitioning, gHyPart}.

This work explores the \textbf{GPU implementation} of an algorithm for \textbf{directed hypergraph partitioning} under partition size and incidence constraints.
Each partition is limited in the number of nodes it contains and in the number of its distinct inbound hyperedges.
The goal of partitioning is to minimize the connectivity, the total weight of cuts induced by hyperedges between partitions.

This constrained formulation emerges across many fields and at scales nearing millions of nodes.
Notable instances are the mapping of Spiking Neural Networks (SNNs) to neuromorphic hardware \cite{AxonFlow, MappingVeryLargeSNNtoNHW}, VLSI/FPGA design under limited I/O budgets \cite{hMETIS_vlsi, AdvancesInHypergraphPartitioning}, chiplet-based architectures \cite{ChipletsCombinatorics}, and sparse matrix kernels \cite{gHyPart}.

A few GPU-based hypergraph partitioners \cite{HyperG, gHyPart} already exist, but focus on balanced $k$-way partitioning, a different set of constraints. 
Still, they demonstrate speedups upwards of $100\times$ over their CPU counterparts \cite{hMETIS_k_way, KaHyPar}.
In doing so, they already highlight challenges in designing algorithms for concurrent decision-making while keeping them GPU-friendly.
Crucial subjects being data access locality, integrating warp primitives, and workload balance.

The introduction of incidence constraints further exacerbates these hurdles.
Tracking distinct inbound connections requires continuous set unions and deduplication.
Then, knowing the validity state of a partition before and after moving a node involves many membership tests on those sets.
Furthermore, these operations are often nested inside neighborhood traversals, already the most costly visit of a hypergraph.
As such, everything, from data structures to kernels, must be bent around these constraints.

Hereafter, we present a new multi-level hypergraph partitioning scheme adapted to our constraints.
To the best of our knowledge, this is the first work to address such a variation of the problem.
We discuss the fundamental bottlenecks of its implementation on GPU, followed by algorithm designs circumventing them.
Key points include the materialization of neighborhoods in memory and sparse, event-based constraint tracking.

\begin{figure}[t]
    \centering
    \includegraphics[width=1.0\columnwidth]{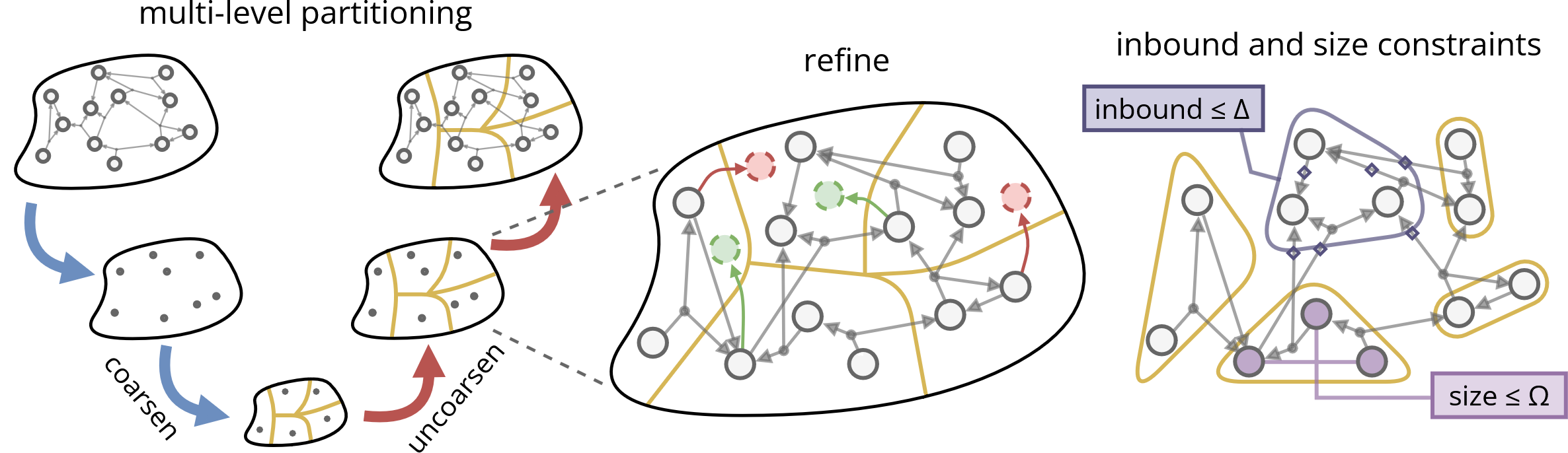}
    \vspace{-12pt}
    \caption{Overview of the multi-level hypergraph partitioning scheme.}
    \vspace{-4pt}
    \label{fig:overview}
\end{figure}

\section{Preliminaries}\label{sec:preliminaries}

\subsection{Hypergraph Partitioning Model}\label{subsec:hgraph_part_model}

Hypergraphs (\hgraphs) are a generalization of graphs where edges -- now hyperedges (\hedges) -- can connect more than two nodes.
A weighted directed hypergraph $G(N, E, \omega)$ consists of a set $N$ of nodes and a set $E$ of \hedges.
Each \hedge $e \in E$ contains a subset of nodes (\hspace{-0.1pt}pins\hspace{-0.1pt}) $e \hspace{-0.4pt}\subseteq\hspace{-0.4pt} N$, with $src(\hspace{-0.3pt}e\hspace{-0.3pt})$ being the \hedge's sources and $dst(\hspace{-0.3pt}e\hspace{-0.3pt})$ its destinations.
The function $\omega : E \rightarrow \mathbb{R}$ assigns a weight to each \hedge.
In addition, we define a node $n$'s incidence sets $in(n) = \{e \in E \sothat n \in dst(e)\}$, $out(n) = \{e \in E \sothat n \in src(e)\}$, $\mathcal{I}(n) = in(n) \cup out(n)$.
Finally, a node $n$'s neighbors are $\mathcal{N}(n) = \{m \in e \sothat e \in \mathcal{I}(n)\}$.

A partitioning of $G$ is a set $P \subset \mathcal{P}(N)$ of pairwise disjoint non-empty subsets -- partitions -- of its nodes such that $\bigcup_{p \in P} p = N$.
With $\mathcal{P}(\cdot)$ denoting the power set.
Equivalently, it can be defined by a $\rho : N \rightarrow P$ such that $\rho(n) = p$ iff $n \in p$.

Our constraints pose hard limits on the number of nodes and of distinct inbound \hedges per partition.
Let $\Omega$ be the maximum size of a partition, $\forall p \in P, \: \abs{p} \leq \Omega$.
Let $\Delta$ be the maximum number of allowed distinct inbound \hedges to a partition, $\abs{\bigcup_{n \in p} in(n)} \leq \Delta$.

Our optimization metric is the \textbf{connectivity}, the weighted number of cuts on each \hedge:
\begin{equation}\label{eq:connectivity}
    Conn_G(\rho) = \textstyle\sum_{e \in E} \omega(e) \cdot (\abs{\{\rho(n) \sothat n \in e\}} - 1) \text{.}
\end{equation}
The partitioning goal is to minimize connectivity within constraints.

While our present focus is on limited distinct inbound \hedges, the presented methodology remains trivially generalizable to constraints on distinct outbound or distinct incident \hedges.

\subsection{The Multi-Level Partitioning Scheme}\label{subsec:multi_level_scheme}

The staple approach to \hgraph partitioning is the multi-level scheme \cite{hMETIS_k_way, KaHyPar} with Fiduccia–Mattheyses refinement \cite{FMpartitioning}, shown in Fig.~\ref{fig:overview}.
It first involves a coarsening phase, progressively clustering together nodes that participate in similar sets of \hedges until further aggregation would violate constraints.
The resulting clusters define an initial partitioning, which is then refined by uncoarsening the hypergraph and evaluating node moves between partitions.

We implement a slight revision of the classic balanced $k$-way version of the method, which relied on a slow, exact algorithm for the initial partitioning of coarse nodes.
With no need for balance, the coarsening routine's goal of maximum \hedge overlap becomes the dual of minimum connectivity.
Moreover, with no $k$ partitions requirement, the least cuts cost is found when a near-minimal number of partitions is formed.
Hence, we let the coarsest level's nodes define partitions, with coarsening halting upon reaching $\lceil \nicefrac{\abs{N}}{\Omega} \rceil$ nodes or being unable to further unify clusters.
This removes the need for the slow initial partitioning step.

While effective, the multi-level scheme often incurs high computational costs.
Namely, during coarsening, clusters must be selected over large and irregular neighborhoods \cite{AcceleratedCoarseningProcedure}.
Whereas for refinement, a subset of improving moves must be extracted from many that mutually interfere \cite{HyperG}.
Now, both such existing bottlenecks are also where inbound constraints checks need to be performed.

\subsection{GPU Model and Data Structures}\label{subsec:gpu_model_data_structures}

Present terminology hinges on CUDA, our chosen API for general-purpose processing on GPU.
Architecturally, a GPU comprises several streaming multiprocessors, each handling several threads grouped in blocks.
Internally, a multiprocessor breaks a block into warps, sets of 32 threads that undergo SIMD execution and can share data through shuffles.
Threads have access to a limited number of registers, while blocks can allocate a small amount of shared memory, a scratchpad seen by all their threads.
Any other data resides in global memory, backed by VRAM.
The result is a hierarchical parallelism model, spanning blocks, warps, and threads.


Two concerns central to GPU programming are warp divergence and memory access coalescing.
To address both, we maintain a compressed sparse memory representation of \hgraphs, their inner structure always traversed by entire warps.

An \hgraph is primarily described by sets of sets, namely $E$, $in(n)$, $out(n)$.
The memory representation of such two-level structures in compressed sparse form involves two arrays.
A segmented data array stores contiguously each linearized inner set.
Then, an array of offsets maps inner set ids to their data's starting position in the previous array.
If now one or few warps handle each segment, they will see fully coalesced accesses and little divergence.
Nodes and \hedges alike are identified by unsigned integers, their ids constituting all atoms inside sets.
With ids being a zero-based range, they double as indices in the offsets array.
In addition, ids enforce a total order over nodes $\prec_{id}N$ and \hedges $\prec_{id}E$.

\section{Coarsening}\label{sec:coarsening}

\subsection{Algorithms Overview}\label{subsec:coarsening_algorithms_overview}

Coarsening starts with the construction of mutually exclusive pairs of nodes, which is carried out in two steps.
First, each node selects among its neighbors the most suitable pairing candidate.
Then, coarse nodes are determined by a maximum-weight matching computed over candidate pairs.

A node $n$'s candidate is the neighbor it is connected to with the highest total weight.
In addition, a candidate must be valid, i.e. it and its node must lead to a valid cluster within constraints.
Candidate selection involves visiting $n$'s incident \hedges and their pins, accumulating for each neighbor the total weight of \hedges it appears in.
This is a histogram over $n$'s neighbors:
\begin{equation}\label{eq:neighbors}
    \forall m \in \mathcal{N}(n), \, hist(n, m) = \textstyle\sum_{e \in \mathcal{I}(n) \text{ s.t. } m \in e} \omega(e) \text{ .}
\end{equation}
Finding the best valid candidate takes repeated extractions of $\argmax_{m \in N}hist(n, m)$, followed by constraint checks until a valid $m$ is found.
Checking cluster size $\leq \Omega$ is trivial, while distinct inbound \hedges require computing the union $\abs{in(n) \cup in(m)} \leq \Delta$.
Finally, each node and its candidate form a candidate pair, $pair : N \rightharpoonup N$, $pair(n) = \max_{id}\,\argmax_{m \in N}^{valid}\,hist(n, m)$, inheriting their total connection's weight as $score : N \rightharpoonup \mathbb{R}$, $score(n) = hist(n, pair(n))$.

Candidate pairs and scores form a directed weighted "pairing" graph overlaying the \hgraph.
Choosing mutually exclusive node pairs reduces to a maximum weighted matching problem over said graph \cite{AcceleratedCoarseningProcedure}.
However, we observe that by virtue of every node proposing one candidate pair, the pairing graph is a pseudo-forest.
Additionally, since both neighbor histograms and validity are symmetric, every edge entering a node has score less than or equal to the edge leaving it, $\forall n, m \in N, \, pair(m) = n \Rightarrow score(n) \geq score(m)$.
This implies that along each component's cycle the score is constant.
In particular, by the definition of $pair$ with $\max_{id}$, all cycles have length two.
So, matching admits a near-optimal solution in two traversals of all connected components \cite{ParameterizedAlgorithms}.


Starting from every leaf in the forest, walking upward until a cycle is found, every node places its score over its candidate. 
Every two-cycle forms a match.
Then, a downward walk goes back from the cycles towards each leaf, matching every node with its candidate iff the candidate is still free and the node's score is still the highest one on it.
The nodes handling order does not matter, so long as the path leading to a node has been fully traced before it is visited.
For brevity, we omit the formalization of nodes with no candidate.

Once final mutually exclusive pairs are determined, they become the next level's coarse nodes.
From there, constructing the coarse \hgraph in full involves merging inbound sets between paired nodes and mapping \hedge pins from nodes to clusters.


\subsection{Parallelization Details}\label{subsec:coarsening_parallelization}

\subsubsection{Neighbors Materialization}\label{subsubsec:neighbors_materialization}


Histogram construction during candidate selection allocates one bin per unique neighbor.
However, without knowing unique neighbors a priori, bins must be overallocated.
With per-node neighbors easily exceeding shared memory capacity, construction of the histogram will spill to global memory.
Hence, turning it into a very costly operation, requiring several random, atomic accesses to global memory.

To streamline the histogram pattern, we fully materialize unique neighbors $\mathcal{N}(\cdot)$ in memory once for the initial \hgraph and progressively update them while coarsening.

This doesn’t alter the histogram’s asymptotic complexity: an iteration over neighbors is still required.
Nevertheless, it brings several advantages.
It offsets the repeated cost of deduplicating neighbors to a single, upfront construction.
When moving down one level, coarse neighbors are computed from existing ones, progressively diluting duplicates.
Deduplication of neighbors alone, rather than bins with weights, occupies exactly half the memory.
Ultimately, the one-time overhead of initial neighbors construction is amortized over all coarsening levels, and histograms are duplicates-free.

Materialized neighbors too follow the compressed sparse format from Sec.~\ref{subsec:gpu_model_data_structures}.
Additionally, they are solely used for coarsening, and their memory can be reclaimed afterwards.

\subsubsection{Candidate Pairs Proposal}\label{subsubsec:candidates_proposal}

\begin{figure}[t]
    \centering
    \vspace{-3pt}
    \includegraphics[width=0.95\columnwidth]{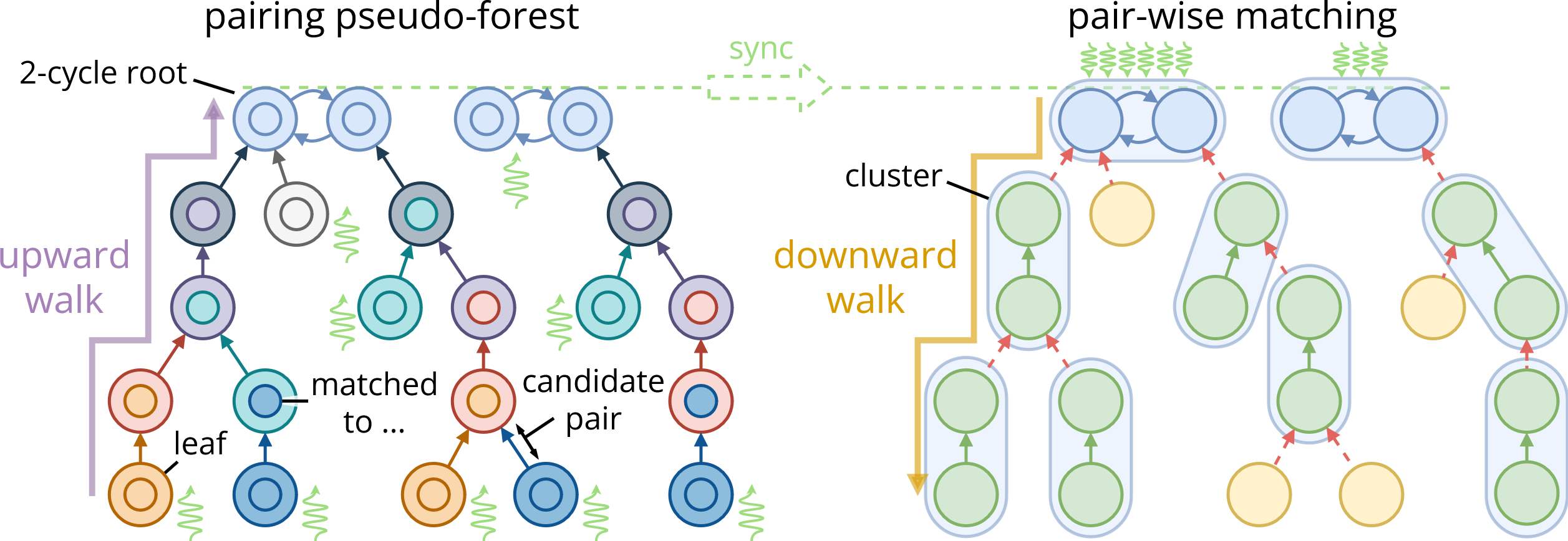}
    \vspace{-3pt}
    \caption{Matching over the two-cycle pseudo-forest.}
    \vspace{-6pt}
    \label{fig:matching}
\end{figure}

With unique neighbors now known, histogram construction is batched over neighbors and happens fully in shared memory.
We assign a warp per node $n \in N$, and let it load a fixed-size batch of $\mathcal{N}(n)$ at once, sorting the resulting bins by key.
The warp iterates $\forall e \in \mathcal{I}(n)$, and as pins are read, a binary search is used to find and increment each bin by $\omega(e)$.
Subsequently, the warp sorts the histogram by value, using neighbor ids as a deterministic tie-breaker.
The current maximum value is extracted and the process repeats until all neighbors have been considered.
Crucially, working only in shared memory allows us to efficiently defer constraint checks until the moment of such extraction, sparing the high cost of validating every neighbor.

Upon extracting the best candidate, cluster size is easily checked by tracking the size of each node while coarsening.
Instead, computing a pair's final inbound set size needs a fast membership test between two nodes' inbound sets.
For this, we require the data array segments of $in(\cdot)$'s compressed sparse form to be sorted.
A warp assigned to node $n$ synchronizes over a single neighbor $m$ at a time.
Threads proceed to read $in(m)$, for each \hedge running a binary search on $in(n)$, reducing with shuffles the count of \hedges not already present in it.
The final total is then added to $\abs{in(n)}$ and checked against constraints.


\subsubsection{Parallel Nodes Matching}\label{subsubsec:parallel_matching}

Following from Sec.~\ref{subsec:coarsening_algorithms_overview}, nodes matching involves two traversals over the pseudo-forest of $pair$ and $score$.
These can occur in parallel with a thread departing from every leaf.
Each node tracks a $match : N \rightarrow N$ indicating its current status, initially $match(n) = n$.
When a thread on node $n$ moves to $pair(n)$, it uses an atomic max operation to set $match(pair(n)) = n$ iff $score(n) >_{id} score(match(pair(n)))$ or $match(pair(n)) = pair(n)$.
Ties are again broken by id.
Two-cycles become "roots" and form a match by default; all threads synchronize upon reaching one.
Then each thread retraces its path back to its leaf, finalizing the match on every node along the way.
If a node $n$ still sees $match(pair(n)) = n$, it locks its match by setting $match(n) = pair(n)$.
Matching nodes will thus point to each other in an alternating pattern, see Fig~\ref{fig:matching}.

As every thread covers the full root-leaf path, they will all take the same decisions.
A minor optimization sees each thread halting on node $n$, before the root, whenever it loses an atomic max, as that already implies $match(n) \neq pair(n)$.

\subsubsection{Coarse Hypergraph Construction}\label{subsubsec:coarse_construction}


With $match$ defining clusters, coarsening finishes by building a new \hgraph $G'(N', E', \omega')$ among them.
With $N' = \{\{n, match(n)\} \sothat n\in N\}$, let $\gamma : N \rightarrow N'$ map nodes to clusters as $\gamma(n) = n'$ iff $n \in n'$.
Then, constructing coarse instances of all two-level structures, $E$, $in(\cdot)$, $out(\cdot)$, $\mathcal{N}(\cdot)$ requires a sequence of map operations through $\gamma$ and set unions.

Both operations involve deduplication, and the final set size is unknown a priori.
Constructing many sets in parallel proceeds in two phases:\;\!\! first, an oversized data array is built while deduplicating, then its segments are packed in compressed form.
Each coarse set is assigned to a block and given both a global memory segment large enough for all its potential elements and the maximum shared memory amount available.
Both memories operate as closed-hashing hash-sets.
Threads collectively read elements from the original set and apply $\gamma$ as needed.
Insertion of each element is first attempted in shared memory, going to global memory only upon a successful insertion or exhausted probe length.
Once all elements are seen, set sizes are known, and a packing operation scatters each oversized segment into its final segment in a new compressed data array.

If the target number of nodes is reached, instead of constructing a new \hgraph, $P \leftarrow N'$ and $\rho \leftarrow \gamma$ define initial partitions.
Otherwise, every step up to this point repeats for $G'$.

\section{Refinement}\label{sec:refinement}

\subsection{Algorithms Overview}\label{subsec:refinement_algorithms_overview}

Local refinement improves a partitioning by selectively moving nodes across partitions.
Choosing suitable moves takes two steps.
First, each node independently selects the partition it would rather belong to, proposing a move.
Then, a subset of moves is found such that, when applied together, they lead to a valid state of lowest possible connectivity.

Each node proposes its move in-isolation.
By Eq.\ref{eq:connectivity}, a cut is only avoided when an \hedge has no pins left in a partition.
So, a favorable move is one that fully disconnects \hedges from the node's current partition while introducing cheaper connections, if any, to the new partition.
To find such moves, a node must count the number of pins each of its incident \hedges owns in every partition.
Let $pins(p, e) = \abs{\{n \in e \sothat n \in p\}}$ be the pins count held by \hedge $e$ in partition $p$.
Then, for every node $n$ and partition $p$:
\begin{equation}\label{eq:refinement_move_selection}
    \begin{aligned}
        & saving(n) = \textstyle\sum_{e \in \mathcal{I}(n) \text{ s.t. } pins(\rho(n), e) =\;\! 1} \omega(e) \text{ ,} \\
        & loss(n, p) = \textstyle\sum_{e \in \mathcal{I}(n) \text{ s.t. } pins(p, e) =\;\! 0} \omega(e) \text{ ,} \\
        & gain(n, p) = saving(n) - loss(n, p) \text{ .}
    \end{aligned}
\end{equation}
Moving node $n$ to partition $p_{d}$ is favorable if $gain(n, p_d) > 0$.
The move with highest gain is proposed by the node.

So-obtained moves have been proposed separately, but to maximize gain, several of them shall be applied at once.
Deciding which moves to apply equates to finding the subset of moves collectively leading to a valid maximum gain partitioning.
However, moves easily interfere, influencing each other's gain and feasibility, making this a problem only solvable in exponential time.
For this reason, we adopt a heuristic from \cite{HyperG}.
We sort moves into a sequence by gain.
In doing so, the problem of deciding which moves to apply reduces to finding the longest subsequence of improving moves landing on a valid state.
To solve it, each move updates its gain in-sequence, i.e. assuming all moves before it already applied.
Analogously, validity over constraints is computed as of every move along the sequence.
A simple filtered maximum extraction then leads~to~the~desired~subsequence.

Moves are subsequently applied before uncoarsening, that is, projecting the partitioning to the next level up.

\subsection{Parallelization Details}\label{subsec:refinement_parallelization}

\begin{figure}[t]
    \centering
    \includegraphics[width=1.0\columnwidth]{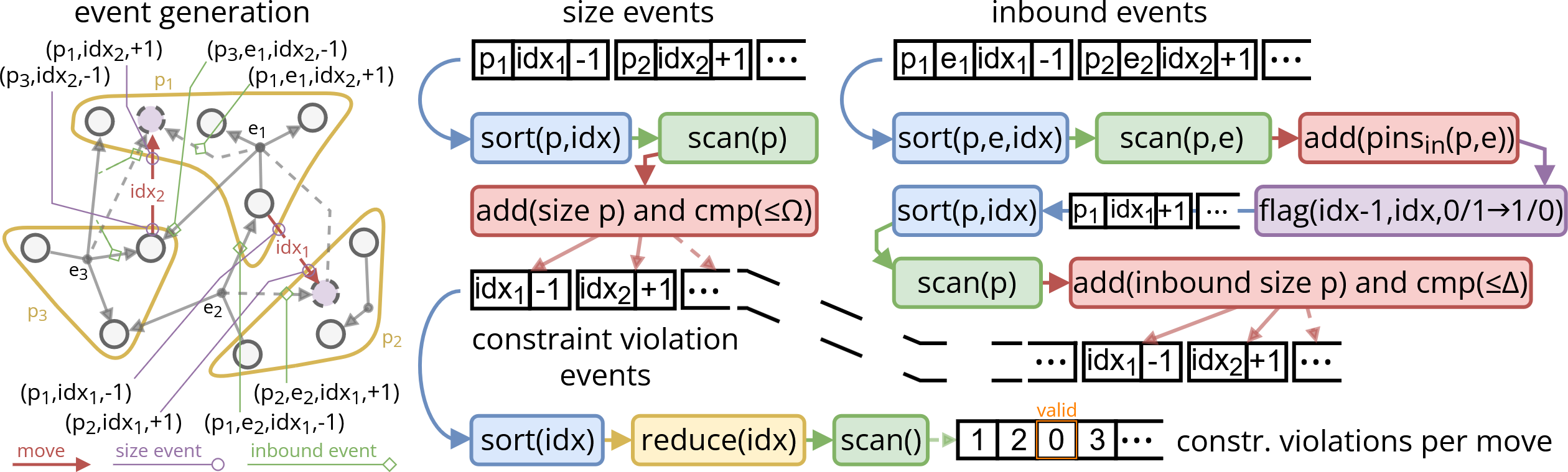}
    \vspace{-12pt}
    \caption{Events-based moves validity check.}
    \vspace{-6pt}
    \label{fig:events_generation}
\end{figure}

\begin{figure*}[t]
    \centering
    \vspace{-6pt}
    \includegraphics[width=\textwidth]{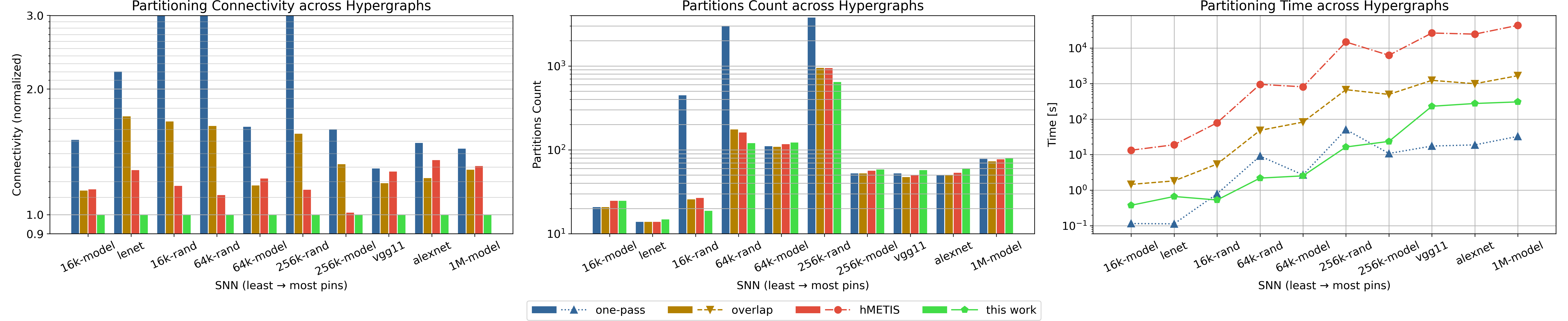}
    \vspace{-16pt}
    \caption{Partitioning results comparison across ten spiking neural network hypergraphs.}
    \vspace{-16pt}
    \label{fig:results}
\end{figure*}

\subsubsection{Counting Pins per Partition}\label{subsubsec:pins_per_partition}


Refinement heavily relies on the count of pins held by every \hedge in each partition, $pins(\cdot, \cdot)$.
With the same values of $pins$ reused multiple times across nodes, they shall be precomputed \cite{HyperG}.
Thus, we prepare them in a matrix $\abs{E} \times \abs{P}$, filled with an iteration of \hedges, mapping every pin to its partition.

It's worth noting that $pins$ can also provide the incidence set's size for a partition as the count of its non-zero entries over \hedges.
So, if we define $pins_{in}(p, e) = \abs{\{n \in dst(e) \sothat n \in p\}}$ as the number of times \hedge $e$ is inbound to a partition $p$.
The distinct inbounds constraint can be written as $\forall p \in P, \: \abs{\{e \in E \sothat pins_{in}(p, e) > 0\}} \leq \Delta$.

Hence, we first prepare $pins$ for moves proposal, then subtract outbound connections from it and use $pins_{in}$ to validate moves.

\subsubsection{Moves Proposal}\label{subsubsec:moves_proposal}

Proposing moves in-isolation starts with each node computing gains over partitions.
Let each node prepare its $saving(\cdot) = 0$ and $loss(\cdot, \cdot) = 0$ entries in shared memory.
Every warp handles a node $n$, currently in partition $\rho(n) = p_s$, with threads iterating $n$'s incident \hedges.
For every \hedge, if $pins(p_s, e) = 1$, $saving(n)$ increments by $\omega(e)$, and for every partition $p_d \neq p_s$, if $pins(p_d, e) = 0$ , $loss(n, p_d)$ increments by $\omega(e)$.
A map-reduce for the maximum gain yields the node's proposed move.
Partitions already of size $\Omega$ are excluded a priori.

With moves sorted by their in-isolation gain, their in-sequence gain must be inferred.
Let a node $n$'s move be $p_s^n \xrightarrow{n} p_d^n$.
Now, a warp iterates over $n$'s incident \hedges and their pins.
For every \hedge $e \in \mathcal{I}(n)$, consider only pins $m \in e$ whose moves precede $n$'s in the sequence; let related moves be $p_s^m \xrightarrow{m} p_d^m$.
Then, two conditions may arise.
If $\abs{\{m \sothat p_d^n = p_s^m\}} - \abs{\{m \sothat p_d^n = p_d^m\}} = pins(p_d^n, e) > 0$ or $\exists \:\! m$ s.t. $p_s^n = p_d^m$ and $pins(p_s^n, e) = 1$, $n$'s gain decreases by $\omega(e)$.
\linebreak
If $\abs{\{m \sothat p_s^n = p_s^m\}} - \abs{\{m \sothat p_s^n = p_d^m\}} = pins(p_s^n, e) - 1 > 0$ or $\exists \:\! m$ s.t. $p_d^n = p_d^m$ and $pins(p_d^n, e) = 0$, $n$'s gain increases by $\omega(e)$.

A final sequence of nodes, sorted by in-isolation gain, and carrying their in-sequence gain, is thus available.

\subsubsection{Longest Valid Improving Subsequence}\label{subsub:improving_subsequence}


Simultaneously validating every move in the sequence requires reconstructing every intermediate state of all partition sizes and, critically, inbound sets.
Materializing all such states is prohibitive; consequently, we handle constraint checks sparsely through "events".
First, each move generates events for every partition size and $pins_{in}$ variation it causes, carrying the variation’s delta as payload.
Next, with a series of parallel patterns over deltas we infer each move's \hbox{validity, see Fig.~\ref{fig:events_generation}}.


Size events are triplets ($p$, $idx$, $\pmslash1$), meaning partition $p$'s size changes by $\pmslash 1$ with the $idx$-th move.
After being sorted by ($p$, $idx$), their deltas are scanned (prefix summed) per $p$;\!\!\; thus, each event stores its partition’s cumulative size variation up to move $idx$.

Inbound set events are quadruplets ($p$, $e$, $idx$, $\pmslash 1$), meaning $pins_{in}(p, e)$ changed by $\pmslash 1$ after move $idx$.
They are first sorted using ($p$, $e$, $idx$) and prefix summed using ($p$, $e$) as keys.
To track inbound set size changes, each delta is combined with its $pins_{in}(p, e)$ and emits a new event ($p$, $idx$, $-1$) when transitioning from $1$ on the event before to $0$ now, or ($p$, $idx$, $+1$) when turning from $0$ to $1$.
Results are again sorted by key ($p$, $idx$) and scanned per $p$, giving $p$'s distinct inbound \hedges count variation as of the $idx$-th move.

With all events sorted by ($p$, $idx$), adding initial set sizes and a comparison with $\Omega$ or $\Delta$ shows if $p$ is valid on move $idx$.
Then, looking at pairs of events for consecutive moves $idx - 1$ and $idx$ tells if move $idx$ was responsible for invalidating or re-validating partition $p$.
This spawns a final sequence of events ($idx$, $\pmslash 1$) every time a partition changes to invalid ($+1$) or valid ($-1$).
When sorted and reduced by $idx$, their prefix sum is the count of active constraint violations after each move.
Only zero-count moves are valid, finding the one of maximum gain gives the subsequence of moves to apply.

\section{Experimental Results}\label{sec:results}

We tested our solution on 10 \hgraphs originating from SNNs and their mapping constraints on neuromorphic hardware \cite{AxonFlow}, see Tab.~\ref{tab:snns}.
The choice of SNNs was driven by their availability across a wide range of sizes and topologies, covering a superset of most other applications.
Our baseline on CPU comprises an implementation of the multi-level scheme in hMETIS adapted to our constraints \cite{hMETIS_k_way, AxonFlow} and a greedy "overlap" heuristic that co-locates nodes based on their incidence sets' overlap \cite{AxonFlow}.
We also include a "one-pass" algorithm that fills partitions with one pass over nodes \cite{MappingVeryLargeSNNtoNHW}, solely imposing constraints.
Our partitioner ran on an A100-SXM4-40GB, while baselines ran sequentially on an EPYC 7453 @ 2.75GHz.

Results are shown in Fig.~\ref{fig:results}.
Our implementation achieves an average speedup of $246\times$ over sequential hMETIS, $15\times$ over the overlap method's single neighborhood traversal, and remaining within $12\times$ of the trivial one-pass method.
These numbers are in line with the SoTA for $k$-way partitioning on GPU \cite{HyperG, gHyPart}.
Moreover, our execution time grows linearly in the number of pins, denoting no significant overhead from the parallel constraints handling logic.

In terms of quality of results, we repeatedly achieve a mean connectivity $0.82\times$ that of hMETIS, $0.71\times$ against "overlap", and $0.09\times$ versus "one-pass".
The achieved number of partitions also reflects these values.
Notably, SNNs named \texttt{-rand} have the most irregular topologies, easily triggering the distinct inbound constraint.
Still, our algorithm settles them with a very limited, yet valid, number of partitions, attesting to its all-round control over constraints.

\slantedssnstable

\vspace{-0.77pt}

\section{Conclusion}\label{sec:conclusion}

Current results show strong promise regarding the scalability of our implementation.
More importantly, they suggest that our handling of alternative partitioning constraints was effective in managing the added complexity.
Presented algorithms are still being improved, with notable avenues being a dynamic programming formulation for exact matching \cite{ParameterizedAlgorithms} and the adaptation of parallelism strategies as the \hgraph coarsens \cite{gHyPart}.
An open-source release is available \cite{AxonCUDARepo}.

\vspace{-0.77pt}


\bibliographystyle{IEEEtran}
\bibliography{biblio_min}

\end{document}